# Self-Enhancing Halite Growth Creates Secondary Porous Networks During $CO_2$ Storage in Saline Aquifers

Mohammad Nooraiepour,[1*] Mohammad Masoudi,[1,2] Hannelore Derluyn,[3] Pascale Senechal,[4] Peter Moonen[3,4] and Helge Hellevang[1]

[1] Department of Geosciences, University of Oslo, P.O. Box 1047 Blindern, 0316, Oslo, Norway.
[2] Applied Geoscience Department, SINTEF Industry, 7465, Trondheim, Norway.
[3] Universite de Pau et des Pays de l'Adour, E2S UPPA, CNRS, LFCR, Pau, France.
[4] Universite de Pau et des Pays de l'Adour, E2S UPPA, CNRS, DMEX, Pau, France.

* Corresponding author: mohammad.nooraiepour@geo.uio.no

**Abstract**

Salt precipitation during $CO_2$ injection into saline aquifers obstructs flow-controlling pore throats and reduces permeability, yet reactive transport models assume salt forms dispersed, non-porous crystals with minimal flow impact. We demonstrate that halite instead creates three-dimensional porous networks with 40±15% internal porosity through self-enhancing growth mechanisms absent from current models. Time-lapse X-ray micro-computed tomography and spectral imaging reveal preferential nucleation at gas-liquid interfaces, where porous hydrophilic aggregates generate capillary suction that draws brine films toward precipitation sites, accelerating growth and expanding reactive surface area in a positive feedback loop. Spectral tomography shows systematic density gradients reflecting two-stage precipitation: dense macrocrystalline cores formed under moderate supersaturation and evaporation, surrounded by microcrystalline overgrowths from rapid late-stage dynamics. These overgrowths create umbrella-like crusts that encapsulate residual brine beneath surface layers. This porous architecture explains why modest porosity reduction causes severe permeability decline, as aggregates preferentially obstruct flow percolation pathways rather than uniformly cementing grains or filling pore space. Five interconnected mechanisms drive self-enhancing growth across nano- to centimeter scales: interface nucleation, secondary porous structure formation, steep concentration gradients, hydrophilic substrate film maintenance, and capillary-driven solute delivery. Our quantitative characterization of internal salt architecture, reactive surface areas, and pore connectivity provides essential parameters for improving predictive models of evaporation-precipitation dynamics in carbon storage, soil salinization, and cultural heritage preservation.

**Significance Statement**

Salt precipitation can impair $CO_2$ storage reservoirs, damage historic monuments, or degrade agricultural soils. Using time-lapse 3D X-ray imaging, we demonstrate that halite builds sponge-like structures with 40% internal porosity rather than simply filling pores. This architecture creates self-reinforcing feedback: porous networks generate capillary forces that draw brine toward precipitation sites, accelerating growth. Critically, these findings resolve why modest porosity reduction causes severe permeability decline, as porous aggregates preferentially obstruct flow-controlling throats at gas-liquid interfaces rather than uniformly filling pore space or grain surfaces. Our quantitative measurements of salt architecture (connectivity, surface area, pore size distributions) provide essential parameters for accurate predictive models, directly improving carbon storage efficiency, agricultural soil management, and cultural heritage preservation strategies.

## Introduction

Global efforts to combat climate change increasingly target greenhouse gas emissions, particularly carbon dioxide ($CO_2$). The European Union aims to achieve carbon neutrality by 2050 (1, 2), with a focus on large-scale deployment of Carbon Capture and Storage (CCS) (3, 4). Meeting these goals requires rapid expansion of $CO_2$ injection facilities across the European Economic Area (4, 5). By 2040, an annual capacity of 250 million tonnes of $CO_2$ is needed, with an interim target of 50 million tonnes by 2030 (3, 6). This requires approximately 100 injection wells by 2030, each injecting 0.5 million tonnes per annum (Mtpa), necessitating one new well every three weeks, which will escalate to one per week by 2040, even with high-performance wells at 1 Mtpa.

Efficient injectivity is critical for large-scale CCS, enabling $CO_2$ flow through reservoir rocks. Injecting supercritical $CO_2$ into geological formations at millions of tonnes annually causes in-situ brine evaporation when the $CO_2$ is undersaturated with water, concentrating dissolved salts (7–9). Salt precipitation occurs when solubility limits are exceeded under reservoir conditions, clogging pores, reducing permeability, and increasing injection pressure (10–15). In high-salinity aquifers, these effects impair $CO_2$ storage performance, injectivity, and containment integrity (16–22). Economically, salt precipitation incurs significant costs through remedial interventions or injection cessation (23–25).

In hypersaline aquifers, $CO_2$-induced salt precipitation intensifies due to complex physicochemical fluid-rock interactions. Globally, several geological carbon storage (GCS) reservoirs exhibit high salinity: Snøhvit (Barents Sea, Norway) exceeds 100,000 parts per million (ppm), Endurance (North Sea, UK) surpasses 250,000 ppm (19), and Quest (Canada) exceeds 270,000 ppm NaCl equivalent (26, 27). Ketzin (Germany) reaches 230,000 ppm total dissolved solids (TDS) (12, 28); Illinois Basin-Decatur (USA) approximates 200,000 ppm TDS (29, 30); Aquistore (Saskatchewan, Canada) reaches 330,000 ppm TDS (11); and Greensand (Denmark) potentially exceeds 100,000 ppm TDS (31). These high-salinity settings challenge $CO_2$ storage operations while remaining critical for global emission reduction efforts.

Salt crystallization in porous media extends beyond carbon sequestration, affecting diverse environmental and engineering domains (32, 33). In soil mechanics, it alters structure and fertility through salinization (33, 34). In agriculture, salt accumulation reduces crop yields and soil health, necessitating intensive management strategies (33). In food processing, it modifies drying kinetics in saturated media, affecting product quality and stability (35, 36). In construction and cultural heritage, salt crystallization via efflorescence (surface deposits) or subflorescence (subsurface precipitation) damages building materials through mechanisms sourced from rainwater, groundwater, marine aerosols, or de-icing salts (37–41), which are detrimental to both aesthetics and mechanical integrity. Environmental cycles, such as wetting and drying, amplify material deterioration (36–38), underscoring the need for a mechanistic understanding to inform preservation strategies.

Prior studies on salt precipitation, primarily focused on surface efflorescence in building materials and soils, have identified capillary pumping, creeping behavior, the efflorescence formation, growth morphologies, and self-amplifying mechanism as key factors during evaporation-crystallization-precipitation processes under atmospheric conditions at open boundaries (33, 37, 42–58). However, these studies did not explore precipitation under the advective-diffusive-reactive flow conditions and confined geometries characteristic of GCS. In GCS, particularly in the near-wellbore region following two-phase viscous displacement under injection conditions, evaporation fronts migrate through three-dimensional (3D) pore networks rather than forming at exposed surfaces. Furthermore, reactive transport models for GCS and existing clogging models lack experimental characterization of the internal architecture of salt precipitates and their evolution under a sustained brine supply in confined porous media.

Although experimental, numerical, and theoretical studies of carbon sequestration in saline aquifers have focused on predicting the quantity and spatial distribution of salt precipitation (7–9), less attention has been devoted to the fundamental physics governing precipitation mechanisms, growth architectures, and interfacial dynamics near the evaporation front. These studies often overlook the critical roles of in-situ brine availability and 3D aggregate structures in controlling crystallization kinetics and aggregation morphology through sustained solute transport.

To address these gaps, we investigate $CO_2$-induced salt precipitation in porous media with sustained brine availability, using time-lapse X-ray micro-computed tomography, spectral imaging, and deep learning-assisted image analysis to track crystallization dynamics from early precipitation through complete drying. Contrary to the assumption that halite precipitates as dispersed, non-porous micro-aggregates, we demonstrate that salt forms secondary porous networks with considerable internal porosity. These structures preferentially nucleate on existing crystals as secondary substrates (59–62), creating a self-enhancing feedback mechanism. We identify five interconnected mechanisms that drive self-enhancing precipitation and quantify internal salt architecture to enable accurate prediction of flow impairment. Our measurements of pore connectivity, surface area scaling, and density gradients provide the missing parameters in reactive transport models for carbon storage, soil management, and heritage preservation.

## Results

### Interface-preferential growth and porous aggregate formation

Figure 1 shows three consecutive stages (t1–t3) of residual brine saturation and salt precipitation in a porous medium under gravitational influence. Cross-sections in the xz and yz planes reveal vertical brine-salt distribution, while xy profiles detail horizontal distributions at different heights along the flow cell. Subfigures B and C-D present magnified views of the regions marked in blue and yellow in panel A, respectively.

$CO_2$ injection displaces brine through viscous two-phase flow, establishing residual brine saturation. Continued $CO_2$ flow drives evaporation of the trapped brine, with rates controlled by brine-$CO_2$ mutual solubility under experimental conditions. The hydrophilic glass beads and halite surfaces shape the gas-brine interface into distinct menisci (Fig. 1, xy views). The aqueous phase coats beads and crystals, clustering residual brine and enhancing wetting-phase connectivity over centimeter-scale distances via capillary-driven flow. Concentration gradients within these films drive molecular diffusion, redistributing dissolved ions toward evaporation fronts.

Figure 1 (C-D, xz, yz views) shows uneven residual brine distribution after drainage, with greater retention in smaller pores due to higher capillary pressures. After drainage, the nearly immobile aqueous phase forms discrete pools of varying sizes. Unsaturated $CO_2$ injection drives evaporation from these brine pools into the gas stream. Magnified view at t1 (Fig. 1B) reveal euhedral to subhedral halite crystals (>100 μm) alongside smaller aggregating micro-crystals. The observed salt morphology exhibits both crystalline facets and semi-rounded aggregate forms in cross-section (Fig. 1B–D). Two-dimensional microfluidic experiments have documented both crystal morphologies as well (19, 21, 24, 44, 63, 64). This departure from ideal euhedral crystal geometry reflects rapid aggregation of multiple crystallites (as proxy for nucleation events) that grow simultaneously at the interface. The circular or irregular profiles in two-dimensional sections represent porous aggregates where individual microcrystals interconnect to form composite structures (Fig. 1B). Reactive transport models (RTM) for $CO_2$ storage typically assume that halite precipitates as dispersed, micron-sized crystals that minimally obstruct flow pathways. The formation of large crystals challenges the prevailing RTM hypothesis that only dispersed, micron-sized crystals form in 3D porous media, necessitating a fundamental revision of existing models that predict minimal flow obstruction from such precipitates.

Pore-scale dynamics from t1–t3 reflect coupled microscale diffusion and transport processes. $CO_2$ injection displaces brine and establishes evaporative fronts where water transitions into the gas phase. Initially, limited brine-gas interfacial area constrains evaporation rates. However, supersaturated micro-environments develop through local concentration gradients, driving halite nucleation and growth via ion migration to active sites. Expanding brine-gas contact area and sustained evaporation maintain ion supply, promoting crystal growth particularly at supersaturated interfaces (Fig. 1) (59–62). The wettability of the porous medium, evidenced by the observed meniscus curvature indicating hydrophilicity of the glass beads and halite surfaces in Figure 1, controls salt distribution and morphology. This crystallization process has important structural consequences. Continuous evaporation, ion transport, nucleation and growth form new crystals atop or beside existing precipitates, creating secondary porous media. These secondary porous networks, characterized by small pore sizes, exert strong capillary forces that imbibe brine. Simultaneously, hydrophilic substrates maintain continuous liquid films that serve as ion transport pathways, enabling crystallization even in regions distant from bulk brine reservoirs.

Individual halite crystals expand and interconnect across stages t1–t3, forming 3D salt networks that progressively alter pore geometry and clog flow percolation pathways. Comparison of xy profiles i–iii in in Figures 1C and 1D reveals a slight increase in precipitation at t3 compared to t2 around the beads, with enhanced brine imbibition observed at the later stage. In contrast, profiles iv–vi document localized dissolution events (Fig. 1C-D, xy views iv–vi, orange circles), revealing a dynamic feedback loop involving precipitation, dissolution, and capillary redistribution. Local dissolution and reprecipitation cycles reflect fluctuating saturation states under variable evaporation rates that collectively drive morphological evolution and generate capillary pressure gradients. These concentration fluctuations generate Marangoni convection or diffusive flows that further redistribute solutes (65–68). Brine movement through interconnected voids sustains solute redistribution: dissolution occurs in undersaturated regions while supersaturation elsewhere accelerates crystal growth, simultaneously increasing reactive surface area available for further precipitation.

Stage t1 exhibits residual brine primarily confined to discrete regions of the porous medium, restricting evaporation due to limited gas-liquid contact area. The xy sections (Fig. 1) demonstrate that halite crystallization occurs both within bulk brine pools and at gas-brine interfaces along percolation pathways. Regions with elevated evaporation rates develop localized supersaturation that promotes crystal growth even in the presence of abundant aqueous phase.

Evaporation from brine films, held by surface tension around existing crystals, generates highly concentrated aqueous pockets particularly conducive to further growth. Sustaining this dynamic equilibrium requires intricate balance between evaporation, solute transport, and crystallization kinetics. Continuous crystal growth establishes a feedback mechanism that perpetuates brine-salt coexistence: growing crystals expand interfacial area, enhancing evaporation and supersaturation, which in turn drives additional precipitation. Conversely, crystallization and dynamic recrystallization may impede brine flow to the fluid interface, thereby suppressing local evaporation, as detailed in subsequent sections.

Tomography images (Fig. 1A, xz, yz views) show a salt creeping phenomenon (42, 43, 55, 58) extending along the flow cell's inner walls in the z-direction. This vertical salt distribution reflects how cell geometry and surface properties channel the aqueous phase, sustaining halite precipitation adjacent to glass beads. The hydrophilic walls facilitate supersaturated micro-environment formation through preferential evaporation at the brine-gas interface, where surface interactions enhance water loss. This localized supersaturation fosters continuous salt precipitation, forming crystalline networks extending over centimeter scales in the vertical direction.

A striking observation from t1 to t3 (Fig. 1) is the persistent coexistence of brine and precipitated halite, seemingly paradoxical given that undersaturated brine would dissolve existing crystals. One might also expect that supersaturated brine would rapidly precipitate dissolved ions, consuming the liquid phase until local thermodynamic equilibrium with the salt is reached. The sustained coexistence observed in three-dimensional visualizations arises because the liquid phase remains in local thermodynamic equilibrium with the salt with respect to halite. Supersaturation drives further crystal growth, albeit at a reduced rate under low supersaturation conditions. Surface tension retains highly concentrated brine films around crystals; these films serve as ionic reservoirs that perpetuate growth. Capillary effects in the hydrophilic porous medium stabilize these thin, supersaturated films within pore spaces, maintaining elevated local ion concentrations that favor continued precipitation over dissolution.

## Temporal evolution from brine-salt coexistence to complete dry-out

Figure 2 tracks temporal evolution from t1 to t5 at 24-hour intervals, documenting the transition from brine-salt coexistence to complete dry-out. At t1, brine and salt coexist during initial growth; by t5, micro-CT reveals extensive salt accumulation throughout the pore network (Fig. 2A–B). At stages t1 and t3, salt precipitates and grows both within brine-saturated regions and at liquid-gas interfaces (Fig. 2A–B). As the brine phase recedes, halite preferentially forms at these interfaces, consuming residual liquid. Large crystals (>100 μm) ultimately extend into the gas phase and protruding through the gas-liquid interface.

Accelerated evaporation during stages t1–t3 generates micron-sized halite aggregates at menisci, manifesting as heterogeneous salt patches at gas-liquid interfaces (Fig. 2B). Sustained solute supply from adjacent brine pools enables these aggregates to evolve by t5 into disordered, interconnected crystal networks, forming secondary porous structures (Fig. 2B–D). From t4 to t5, forced evaporation in the oven depletes large brine pools, shifting precipitation from large crystal formation to micron-sized aggregates due to rapid brine loss and intensified evaporation rates (21). Loss of brine film continuity during these late stages further accelerates aqueous phase consumption.

During final stages (t4–t5), crystallization continues as long as brine remains available, occurring predominantly on existing salt surfaces that facilitate residual brine migration from bead surfaces. Strong capillary-driven flow draws solutes toward active precipitation sites. The hydrophilic nature of halite promotes ion retention on crystal surfaces, generating increasingly intricate aggregates that progressively obstruct pore throats and flow pathways (Fig. 2B&E, t5). Figure 2C inset highlights the role of residual brine saturation in controlling aggregate morphology. Retained internal brine sustains continued growth at t5, maintaining porous structures with elevated liquid saturation (~10±2% residual brine content) even after oven drying. Prolonged oven drying eventually evaporates this internal brine, yielding fully dried salt aggregates with minimal (less than <3%) additional crystallization. Complete dry-out during $CO_2$ injection into saline aquifers is challenging and likely unattainable, as salt surfaces retain residual brine continuity (49). This is critical for reactive transport modeling of $CO_2$-induced salt precipitation, as models often assume complete dry-out, potentially underestimating salt growth impacts.

Figures 2D and 1B illustrate brine-salt interactions around a glass bead. Segmented images show a continuous liquid film coating the bead surface, overlain by salt precipitates. Direct salt-bead contact indicates preferential nucleation at low-energy interfacial sites. The 3D salt structure contains internal voids and inter-crystallite porosity generated by aggregating precipitate patches, sustaining dynamic equilibrium by maintaining brine access to active growth sites (Fig. 2D).

Segmented phase volumes quantify brine and salt evolution from t1 to t5 (Fig. 2G). Glass beads occupy 60±2% of total internal volume of the flow cell, with the remainder distributed among brine, halite, and voids. Kernel density estimates (KDE; see Methods) reveal phase volume distributions and spatial variability (shaded regions: ±5% confidence intervals from subvolume analysis). Salt volume increases progressively from t1 to t4, then stabilizes at approximately 10% by t5. Porosity remains near 30% in the final stage. Rising brine volume during intermediate stages (t2–t4) reflects capillary-driven flow supplying solutes to precipitation sites. Limited salt growth between t4 and t5 indicates depletion of accessible brine reservoirs, with void space increasing as water saturation approaches zero.

Tomographic and segmented images of the pore space reveal distinct spatial patterns in halite precipitation (Fig. 2). As Figure 2F shows the uppermost layers exhibit umbrella-like crusts formed by progressive accumulation at the gas-brine interface where evaporation is most intense (as indicated by label 1). These crusts trap residual brine within underlying pore spaces, simultaneously sustaining crystal growth while inhibiting further evaporation and thereby prolonging the drying process (label 2). The observed spatial patterning resembles the coffee-stain effect (69, 70): faster evaporation at droplet edges drives outward radial flows that transport solutes toward the periphery, creating dense peripheral deposits with sparse interior accumulation. Beneath these crusts, distinct fine-grained cubic halite crystals (typically 1-10 μm) indicate localized growth under moderate supersaturation (label 3). Halite nucleates at the brine-gas interface and grows into the gas phase, forming voids between salt aggregates and bead surfaces. Crystals preferentially expand into the less viscous gas environment (label 4), leaving mobile interstitial brine that redistributes through capillary action.

Internal aggregate architecture consists of porous networks of sub-micron halite crystals interspersed with voids and brine-coated surfaces (Fig. 2C&F). These structures entrap isolated brine pockets, reflecting dynamic equilibrium between crystallization and liquid retention. Continuous brine supply fosters secondary aggregate growth, forming interconnected porous networks that control fluid flow and ion transport pathways (label 5 in Fig. 2F).

The observed crystallization patterns significantly impact pore space characteristics, particularly the porosity-permeability relationship. Crystal growth obstructs pore throats and connected percolation pathways, forming banded aggregate structures (Fig. 2E) that create localized blockages within the pore network (Fig. 2F). While these aggregates cause only moderate porosity reduction, they produce disproportionately large permeability decreases as observed in similar systems (18, 71). Although direct permeability measurements were not performed in this study, the observed pore throat obstruction patterns suggest comparable behavior. This disparity raises a critical question addressed in the following section: do these initially porous aggregates evolve into non-porous forms, or do they retain sub-resolution intercrystalline porosity through densification or microstructural reorganization?

**Bimodal density distribution within halite precipitates**

Spectral X-ray micro-computed tomography (Sp-CT) reveals compositional and density heterogeneity within halite aggregates that conventional micro-CT cannot resolve. We analyzed attenuation coefficients (μ) at low-energy (21–31 keV, emphasizing compositional contrast) and high-energy (51–61 keV, emphasizing density contrast) to differentiate halite's dense macrocrystalline cores from porous microcrystalline overgrowths. Figure 3A presents a schematic of the energy-dependent linear attenuation coefficient as a function of energy, with corresponding spatial distributions of the μ values averaged over the 20-160keV range shown in Figure 3B. A heat map of the average μ-value in the range 21-31keV versus the average μ-value in the range 51-61keV for each pixel in the image (Fig. 3C) features four distinct clusters that can be associated with fluid (region 1), glass flow cell (region 2), glass beads (region 3), and halite aggregates (region 4), as segmented in Figure 3D. Within the halite cluster, μ values reflect substantial density heterogeneity, spanning low-density porous precipitates to dense, compact aggregates. Segmented profiles (Figs. 3E–F) reveal halite nucleating on glass bead surfaces and extending into adjacent pore spaces, with spatial density variations reflecting internal porosity and intercrystalline void spaces.

Detailed analysis of halite aggregates reveals systematic density architecture (Fig. 3C, F): dense macrocrystalline cores are surrounded by lower-density overgrowths of porous microcrystallites, particularly pronounced at the top and bottom aggregate surfaces. This bimodal density distribution indicates a two-stage precipitation sequence. Dense cores formed under moderate supersaturation during slower evaporation, while porous overgrowths developed during rapid evaporation in the final drying stage when supersaturation increased sharply. This spatial segregation, characterized by dense interiors and porous peripheries, mirrors the coffee-stain effect described earlier (46, 47), where capillary flows transport solutes toward evaporation fronts at aggregate surfaces.

The umbrella-like crystal morphologies observed in our system (Fig. 3F) exhibit particularly low densities at growth tips extending into the brine-gas interface and at junctions where neighboring aggregates merge. These structures contain substantial internal pore volumes (porosity quantified below) filled with loosely consolidated microcrystallites. The combination of narrow intercrystal gaps and proximity to high-evaporation interfaces results in open, poorly consolidated crystalline extensions rather than compact, pore-filling structures.

**Three-dimensional internal architecture and pore network quantification**

High-resolution scanning electron microscopy (Fig. 4A–B) at 50 nm imaging resolution reveals two distinct halite surface textures within the sample. Figure 4A shows a smooth, dense texture with minimal visible porosity (non-porous or nano-porous at the resolution limit), while Figure 4B displays a micro-porous texture with interconnected crystals and channels. The non-porous texture (Fig. 4A) exhibits compact surfaces with few resolvable voids. The micro-porous texture (Fig. 4B) contains interconnected, brick-like pore structures with diameters of 5–40 µm measured on SEM micrographs, indicating sustained pore connectivity during crystal growth.

To quantify internal porosity within halite aggregates and textural architecture, we reconstructed 3D structures of extracted crusts and aggregate interiors using X-ray micro-tomography (Fig. 4C-E). These crusts (color-coded in violet on Fig. 4C) correspond to the umbrella-like structures identified in Figures 2-3. Figure 4C presents 3D renderings depicting the precipitation texture, growth morphology, and multiple layers of crust and halite aggregate interiors. The 2D profiles (xz and xy views, Fig. 4C) illustrate the umbrella-like crust layer encapsulating the interior while growing upward into the gas stream and developing intercrystalline porosity, which is clearly visible in these images. The figure also highlights feeding columns through which brine was transported via capillary forces to the top layer, where evaporation-precipitation conditions were comparatively more favorable. Notably, the internal structure reveals cubic halite crystals, porosity within aggregates, and their role in forming a secondary substrate for the interior salt crust.

Segmentation into salt and void phases reveals that surface crusts, which appear as non-porous textures in 2D sections from top view, contain approximately 40% internal porosity in 3D. Pore network modeling via connected component analysis quantifies pore structure (Fig. 4D): diameters range from 20–200 µm (mean 70 µm, 95th percentile 100 µm), coordination numbers range from 1–26 (mean 5), and pore surface areas are predominantly <0.12 mm² (mean 0.04 mm²). These parameters enable capillary pressure calculations and provide reactive surface area estimates for reactive transport modeling of $CO_2$-induced precipitation.

Halite aggregate interiors exhibit highly porous architecture (up to 62% porosity) with diverse crystal morphologies (Fig. 4E–F). A cylindrical section segmented into halite and void phases reveals internal voids and layered branching patterns in 2D profiles. For instance, Figure 4E (xy view) reveals cubic halite crystals comprising columnar structures with lateral protrusions and internal void spaces. These form euhedral to subhedral crystals that extend into surrounding pore spaces and interconnect with both basal and crustal layers. The resulting internal porosity demonstrates strong connectivity in horizontal and vertical directions, establishing a pervasive 3D network (Fig. 4E). An exemplary section is labeled in Figure 4F with measured dimensions in millimeters, characterizing the base layer thickness, structure of fully developed euhedral crystals, and crosscutting crust layers incorporating solute feeding columns. This depiction elucidates the architecture of umbrella-like crusts and aggregate interiors, highlighting their multiscale size relationships within a unified precipitation-growth environment in reactive porous environments.

Pore size distribution analysis (Fig. 4G) shows that most pores have diameters <70 µm and volumes <0.003 mm³, clustering in the lower range of the distribution. Larger pores correspond to surface-connected voids along aggregate boundaries. Multi-scale porosity analysis across hierarchical volumes ranging from 0.001 to 0.015 mm³ (Fig. 4G) reveals scale-dependent heterogeneity: internal porosity averages 35±15%, with greater variability in larger sample volumes (>0.75 normalized volume) compared to smaller subvolumes (<0.2 normalized volume) (Fig. 4F). Best-fit trends and KDE indicate that larger analysis volumes capture greater structural heterogeneity, while smaller volumes sample more uniform local porosity. This scale-dependent behavior reflects the hierarchical organization of pore networks within salt aggregates.

**Discussion and Conclusions**

Our findings challenge the prevailing assumption that halite precipitates as dispersed, non-porous aggregates, with only cementing behavior on the grain surfaces with minimal impact on fluid percolation pathways during $CO_2$ injection in saline aquifers. Instead, we demonstrate that salt inside the porous medium forms three-dimensional secondary porous networks with 40±15% internal porosity through nucleation at gas-brine interfaces, growth within both aqueous and gas phases during brine-salt coexistence, and aggregation into complex morphologies. These structures create self-enhancing (a.k.a. self-amplifying) feedback mechanisms (24, 55–57, 64, 72), analogous to efflorescence liquid capillary pumping effect (51–54), that accelerate precipitation beyond predictions of current reactive transport models. Although salt aggregation and self-amplifying behavior have been observed in prior surface efflorescence studies, these investigations have not quantified the internal architecture in three dimensions inside porous media relevant to GCS. Our work addresses this gap by providing the first 3D characterization of salt growth architecture. It also demonstrates that self-enhancing growth in confined porous media arises specifically from the coupling of porous network formation, substrate characteristics, and capillary-driven transport. Such interactions cannot be captured through two-dimensional observations or open-surface evaporation systems. This self-enhancing growth emerges from coupled processes operating across multiple spatial scales that collectively sustain supersaturation and expand reactive surface area, perpetuating precipitation even as bulk brine recedes.

The self-enhancing mechanism arises from hierarchical coupling across three spatial scales, as described below. At the microscale (nm-μm), surface chemistry governs: hydrophilic substrates (mechanism D) stabilize brine films that maintain solute transport pathways even after bulk liquid drainage. At the mesoscale (μm-mm), architectural effects dominate: porous salt networks (mechanism B) generate strong capillary pressures driving film flow (mechanism E), while expanding gas-liquid interfaces increase reactive surface area for nucleation (mechanism A). At the macroscale (mm-cm), concentration fields control behavior: steep gradients (mechanism C) develop from evaporation fronts toward brine reservoirs, sustaining long-range solute delivery. These scales couple bidirectionally through feedback: porous precipitation (meso) amplifies capillary forces (micro) that steepen concentration gradients (macro) by enhancing evaporation, which in turn accelerates precipitation (meso), completing the self-reinforcing feedback cycle.

Scanning electron microscopy reveals that surface texture correlates with local precipitation rate. Non-porous textures (Fig. 4A) form under rapid crystallization when high supersaturation and evaporation produce compact surfaces. Micro-porous textures (Fig. 4B, individual pores 5-40 μm) develop under slower growth conditions where moderate supersaturation enables brine film maintenance and sustained solute transport through interconnected channels. This textural distinction within individual aggregates (Fig. 4C-E) reflects temporal evolution as evaporation intensifies during drying, consistent with the bimodal density distribution observed in spectral tomography (Fig. 3).

The observed density and porosity variations reflect complex interplay between nucleation kinetics and crystal growth rates under varying supersaturation. Counterintuitively, high supersaturation drives rapid, kinetically limited precipitation that produces porous microcrystalline aggregates with extensive intercrystalline voids. Under these conditions, abundant nucleation occurs, but individual crystals grow rapidly without sufficient time for structural consolidation. Consequently, they coalesce into a seemingly non-porous surface texture, as observed in 2D images. Conversely, lower supersaturation promotes slower, thermodynamically controlled growth, enabling brine redistribution through capillary forces, ultimately yielding dense, well-consolidated macrocrystals. These morphologies correspond to "patchy" (open) and "crusty" (closed) structures in the efflorescence literature (54, 73, 74). This counterintuitive supersaturation-porosity relationship has been documented in other evaporite systems but is not incorporated into reactive transport models for carbon sequestration.

Five interconnected mechanisms drive halite's self-enhancing growth:

**(A) Nucleation at expanding gas-liquid interfaces.** Evaporation progressively shifts the gas-liquid interface from larger to smaller pores, expanding interfacial area available for precipitation. Reduced interfacial tension and elevated solute concentrations at gas-liquid interfaces (relative to solid substrates) promote ion aggregation into stable nuclei (59, 75, 76), creating finely-distributed aggregated crystals. This self-perpetuating cycle—evaporation driving nucleation, nascent nuclei attracting additional ions—establishes conditions for expansive precipitation.

**(B) Formation of secondary porous structures.** Halite precipitation generates porous networks through crystal interconnections on secondary substrates. Small pores (<200 μm to potentially 100s of nanometers) generate strong capillary pressures amplified by high meniscus curvature and steep evaporation gradients (77, 78). These structures sustain brine influx through capillary suction, forming a positive feedback loop: porous precipitates draw more brine, enabling additional precipitation that further increases porosity and capillary strength.

**(C) Steep concentration gradients at evaporation fronts.** Evaporation creates concentration gradients that peak at the drying interface (e.g., near-wellbore regions) and decline with distance from the evaporation front (7, 18, 33). Governed by Fick's laws, diffusive ion flux maintains supersaturation at precipitation sites (79). Higher evaporation rates steepen these gradients, with Marangoni convection enhancing solute transport (65–68).

**(D) Hydrophilic substrate wettability.** Hydrophilic surfaces (glass beads, halite crystals) stabilize thin brine films through low contact angles, enabling solute transport to evaporation fronts even after bulk brine recedes (21, 80, 81). Halite's hygroscopicity further enhances moisture retention, maintaining local supersaturation (82, 83). This substrate-driven film stability is particularly important in dynamic near-wellbore environments experiencing localized drying.

**(E) Capillary-induced brine film flow.** Capillary forces drive brine film flow through thin liquid layers, operating from micrometer-scale pores to centimeter-scale networks (84–86). This transport links distant brine reservoirs to active evaporation fronts near injection wells. Film flow efficacy depends on pore connectivity, hydraulic conductivity, and evaporation site distribution, with hydrophilic halite surfaces enhancing capillary pull. When advection dominates diffusion (Péclet number $Pe >> 1$), rapid solute delivery triggers nucleation and precipitation (87, 88). The resulting porous precipitates increase connectivity and capillary suction, creating a feedback loop (linked to mechanisms A–D) that sustains growth by continuously supplying supersaturated brine to nucleation sites. This coupling between crystallization and capillary forces dynamically reshapes pore geometry and fluid flow properties.

These five mechanisms operate synergistically across the three spatial scales described above. Microscale film stability (D) enables mesoscale capillary transport (B, E) that maintains macroscale concentration gradients (C), while interface evolution (A) couples all three scales by simultaneously expanding reactive area (meso), steepening concentration fields (macro), and enhancing film-mediated ion delivery (micro).

## Future Directions

Understanding salt precipitation during $CO_2$ injection requires quantifying the dynamic interplay between capillary, viscous, and gravitational forces that govern fluid movement, solute transport, and deposit evolution. Current reactive transport models lack a unified framework to predict when self-enhancing feedback dominates versus when precipitation remains localized, limiting their ability to forecast injectivity impairment and potential containment integrity challenges.

We propose a dimensionless number approach as a future direction to characterize force balance regimes and identify rate-limiting processes. The Capillary number ($Ca = \mu v/\gamma$) distinguishes viscous-dominated displacement from capillary-controlled regime; the Bond number ($Bo = \Delta\rho g l^2/\gamma$) quantifies gravity-driven brine redistribution that sustains near-wellbore solute sources; the Péclet number ($Pe = vl/D$) determines whether advection or diffusion controls solute delivery to precipitation sites; the Damköhler number ($Da = kl/v$) identifies whether crystallization kinetics or transport limits growth rates.

Integrating these metrics with the measured internal porosity, pore connectivity, pore-throat network map, and reactive surface areas would enable reactive transport models to implement and capture self-enhancing mechanisms—specifically, how porous precipitates amplify capillary suction, accelerate solute delivery, and expand reactive surfaces, creating positive feedback absent in current models assuming non-porous aggregates. This multi-physics approach is essential for designing $CO_2$ storage operations that maintain injectivity while ensuring containment security.

## Materials and Methods

### Experimental design

This study investigated pore-scale halite precipitation patterns and growth dynamics within a porous medium of glass beads, driven by brine film movement under the influence of viscous, gravitational, and capillary forces. We used a custom-designed cylindrical glass flow cell with an outer diameter of 7 mm and a height of 100 mm, fused to a capillary tube with an outer diameter of 1.5 mm and a height of 50 mm. The flow cell was packed with glass beads of 2 mm (±50 μm) diameter and positioned vertically. First, it was vacuum-saturated with a 5 M sodium chloride (NaCl) stock solution dissolved in deionized ultrapure water (Milli-Q water) at ambient temperature (19±1°C). $CO_2$ was then injected in the gaseous phase at a pressure of 1 MPa through the bottom port at a flow rate of 5 mL/min, displacing the brine and establishing residual brine saturation through viscous displacement (stage 1 as described in our prior work (21)). Following drainage, dry $CO_2$ flow continued to drive evaporation and initiate salt precipitation.

Salt crystals formed in the top 25 mm of the column during this initial phase were imaged at time t1 (stage 2, as described in (21)). We then monitored spontaneous evaporation under no-flow ambient conditions to monitor brine phase movement and salt growth structure at t2 and t3 at 24-hour intervals. This was followed by accelerated evaporation in an oven at 60±2°C at t4 and t5 at 24-hour intervals, which were imaged the next day after returning to ambient temperature. Micro-CT imaging was conducted after each evaporation-precipitation stage. Halite aggregates on glass beads were extracted post-experiment and imaged separately to enable detailed microstructural analysis.

### Time-lapse X-ray computed tomography (micro-CT)

Time-lapse X-ray computed tomography was performed to track the three-dimensional evolution of brine distribution and halite precipitation from initial crystal formation through complete drying. Imaging was conducted using a Tescan UniTOM XL operating at 60 kV and 14 W. Five temporal stages were captured: t1 (after initial $CO_2$-driven halite precipitation), t2 and t3 (following ambient evaporation at 24-hour intervals), and t4 and t5 (after oven drying at 60°C at 24-hour intervals). Two scanning protocols were followed: full region of interest (ROI) scans (2142 projections, 85 ms exposure time, 5 μm voxel size, ~10 minutes acquisition time) provided entire ROI imaging, while high-resolution zoomed scans (2700 projections, 180 ms exposure time, 2.5 μm voxel size, ~33 minutes acquisition time) captured detailed crystallization features in the active precipitation zone. We reconstructed acquired data using Panthera (v 1.4.3.16, Tescan). All analyses focused on the upper 25 mm region where evaporation-driven precipitation was most active, ensuring consistent comparison across temporal stages.

### Spectral X-ray computed tomography (Sp-CT)

Spectral X-ray computed tomography was performed at the final experimental stage (t5) to characterize compositional and density variations within halite aggregates that cannot be resolved by conventional micro-CT (absorption contrast). Sp-CT exploits the energy-dependent transmission of X-rays through the sample to untangle the effects of material composition and density on the attenuation of the X-ray beam (89). The magnitude of the attenuation μ depends on both atomic number (Z) and physical density (ρ). While density effects remain linear across the energy spectrum, compositional effects vary strongly with photon energy, enabling material differentiation (90). Imaging was performed using a Tescan UniTOM XL Spectral, equipped with a linear cadmium telluride (CdTe) energy-resolving photon-counting detector (307 mm × 0.8 mm active area, 764 × 1 pixel array, 140 channels of 1 keV energy bins from 20-160 keV) at the DMEX Center for X-ray Imaging (Université de Pau et des Pays de l'Adour, France). A total of 800 projection images were acquired over a 360° rotation during a 20-minute scan and reconstructed into a spectral energy-resolved volumetric slice with an isotropic voxel size of 9.5 μm using the Tescan Spectral Suite (v.2.1.0.108). For material characterization, we analyzed attenuation coefficients in two energy windows: low-energy (21-31 keV, emphasizing compositional contrast) and high-energy (51-61 keV, emphasizing density contrast). This dual-energy approach enabled segmentation of materials within the cell based on their distinct attenuation signatures.

**Scanning electron microscopy**

We performed scanning electron microscopy (SEM) with backscattered electron (BSE) and secondary electron (SE) imaging to study the post-experiment surface textures and growth structures at sub-micron resolution. SE imaging provided high-resolution surface topography, while BSE imaging revealed compositional contrast based on atomic number differences. A variable-pressure Hitachi SU5000 FE-SEM (Schottky Field Emission Gun) with a Dual Bruker XFlash system was employed. Specimens were mounted on aluminum stubs and coated with a thin layer of carbon using a Cressington 208C high-vacuum coater prior to imaging. This process ensured electrical conductivity, enhanced image quality, and facilitated thorough topographic examination, while preventing charging artifacts and thermal damage.

**Deep learning-assisted segmentation and quantitative analysis**

Tomographic datasets were processed using Dragonfly software (v2024.1) in conjunction with an in-house Python image processing pipeline. Reconstructed volumes underwent noise reduction using non-local means filtering to preserve edge features while suppressing noises. Phase segmentation was performed using a U-Net convolutional neural network trained on manually annotated slices as ground truth data. This approach differentiated five phases with overlapping grayscale intensities, namely glass flow cell walls, glass beads, halite crystals, brine, and void space, which could not be distinguished using traditional thresholding techniques. The deep learning (DL) network architecture consisted of 5 encoder-decoder levels with 64 initial convolutional filters, trained for 100 epochs using 64×64×1 pixel patches with categorical cross-entropy loss and the Adadelta stochastic optimization algorithm. Data augmentation was incorporated to increase the variety of training samples by a factor of 10 (flipping, rotation ±180°, zoom 0.9-1.1×, shear ±2.0°) to enhance model robustness. The trained DL model was then applied to entire 3D volumes, enabling quantitative analysis of phase distributions, crystallization patterns, and temporal evolution of brine-salt coexistence dynamics throughout the experimental timeline (t1-t5).

Phase volume distributions (Figures 2, 4) were visualized using Kernel Density Estimation (KDE) with Gaussian kernels, providing non-parametric probability densities. KDE probability density clouds reveal data concentration (peaks indicate most probable values) and spatial heterogeneity. Shaded regions represent ±5% confidence intervals calculated from spatial variability across subvolumes. These confidence intervals do not assume normal distributions but rather reflect the actual spatial variability observed in the segmented volumes. Bandwidth selection was optimized to balance smoothing and fidelity to the data distribution, revealing underlying trends while minimizing noise artifacts.

Three-dimensional pore network models of segmented halite structures were extracted and analyzed using OpenPNM (91). Connected component analysis identified individual pore bodies and throats within the secondary porous salt networks. Pore geometric properties, including inscribed diameter, volume, surface area, and coordination number (number of connected neighbors), were computed for each pore body. Multi-region analysis quantified porosity scaling behavior across hierarchical volumes ranging from 0.001 to 1.0 $mm^3$. Pore size distributions and connectivity statistics provided quantitative metrics of internal salt architecture relevant for reactive transport modeling.

**Acknowledgments**

**Funding:** This work was supported by the Norwegian Centennial Chair (NOCC) programme through the project "Understanding Coupled Mineral Dissolution and Precipitation in Reactive Subsurface Environments." This project has received funding from the European Union's Horizon Europe research and innovation programme under grant agreement No 101131765 (EXCITE2) for Transnational Access conducted at UPPA-DMEX, France. H. Derluyn acknowledges support from the European Research Council (ERC) under the European Union's Horizon 2020 research and innovation programme (grant agreement No. 850853). This research was conducted using Dragonfly 3D World, provided by Comet Technologies Canada Inc., Montreal, Canada.

**Competing interests:** The authors declare that they have no competing interests.

**Data availability:** All tomography data will be deposited in the DataverseNO public repository upon publication with an assigned DOI. Raw projection datasets exceed repository size limits but are available from the corresponding author upon request.

## Figures

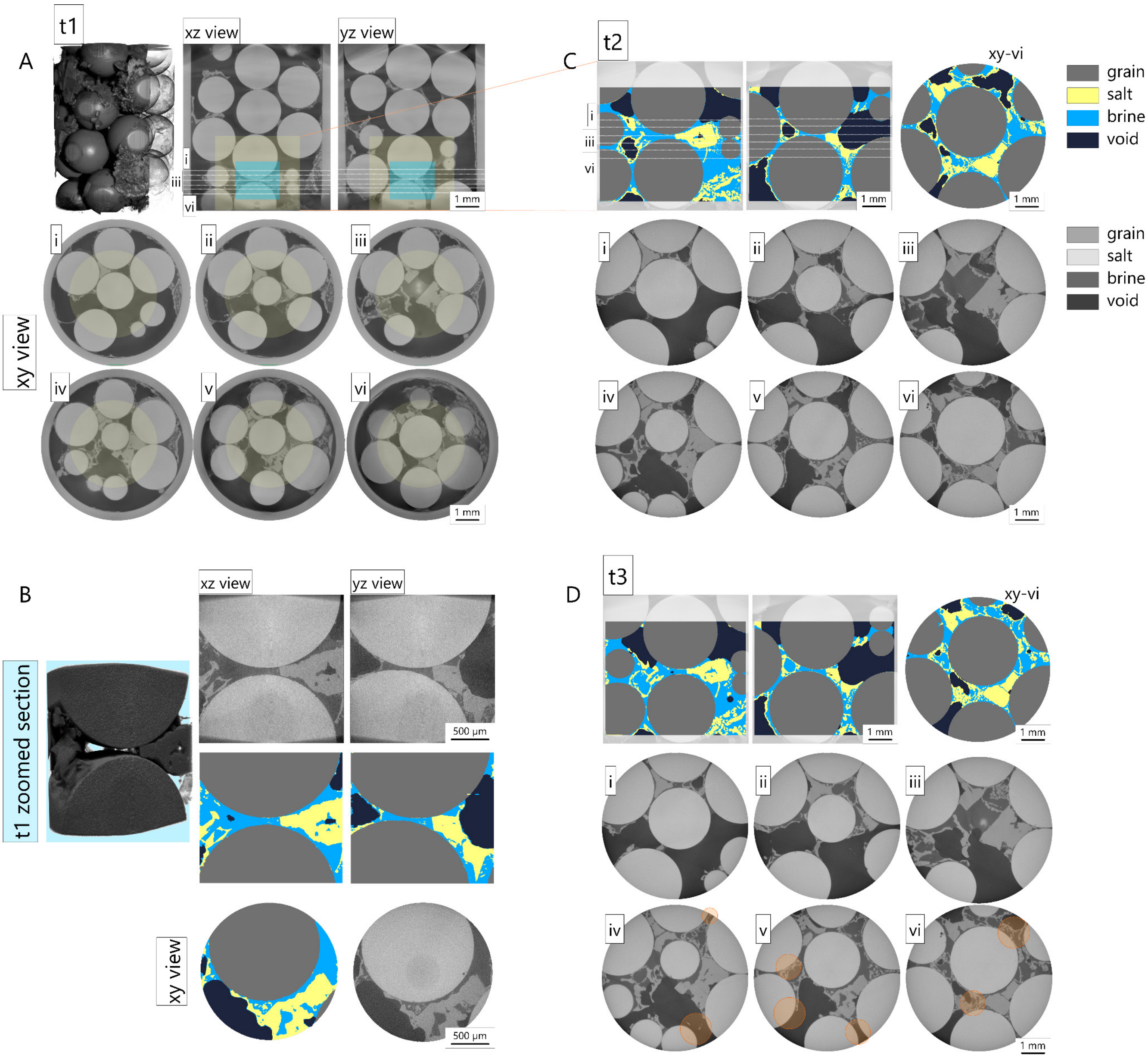

**Figure 1. Interface-preferential salt growth and porous aggregate formation during evaporation-precipitation.** A. Time-lapse micro-computed tomography at stage t1 showing glass bead porous medium following $CO_2$-driven drainage. Magnified regions marked in blue and yellow are detailed in panels B and C–D, respectively. B. High-resolution magnified view at t1 from the blue region showing euhedral to subhedral cubic halite crystals (>100 µm) coexisting with smaller aggregating microcrystals around a glass bead. Rounded profiles in segmented images reflect three-dimensional porous aggregates of interconnected microcrystals rather than isolated euhedral crystals. C–D. Temporal evolution from t2 to t3 in yellow-marked regions on subfigure A. Vertical sections (xz, yz) reveal capillary-driven brine retention patterns; horizontal xy profiles at six heights (locations marked by white dotted lines) demonstrate preferential precipitation at gas-liquid interfaces. The xy horizontal profiles track residual brine and halite distribution dynamics: profiles i–iii show increased salt accumulation and enhanced brine imbibition at t3, while profiles iv–vi (orange circles) document localized dissolution-reprecipitation events, revealing dynamic feedback involving precipitation, dissolution, and capillary redistribution under fluctuating supersaturation.

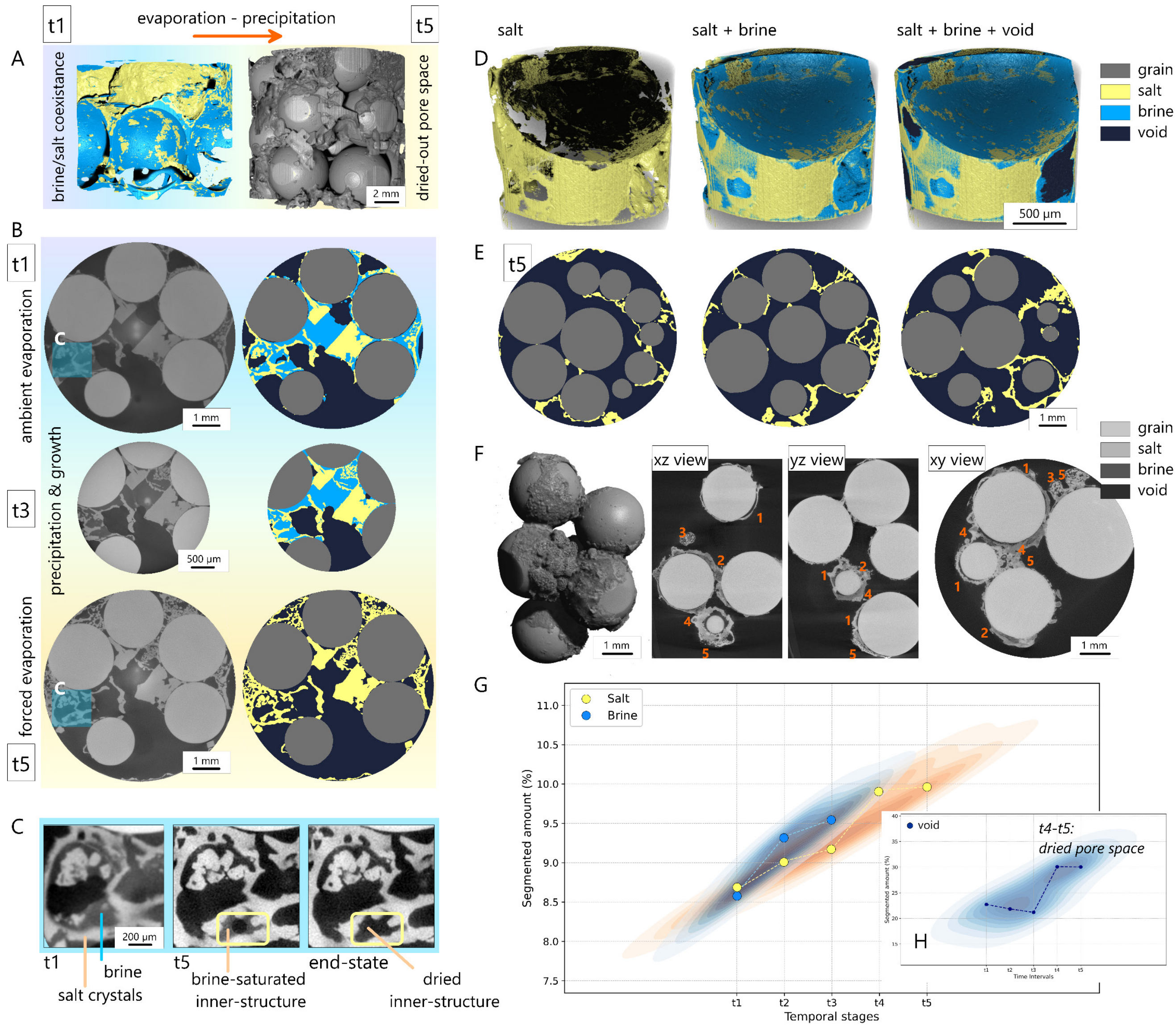

**Figure 2. Temporal evolution and spatial patterning of salt precipitation from brine-salt coexistence to complete dry-out.** A–B. Time-lapse tomography and segmented images tracking halite precipitation across five stages (t1–t5 at 24-hour intervals) from initial brine-salt coexistence to extensive salt accumulation in fully dried pore space. Early stages (t1–t3) show precipitation both within brine-saturated regions and preferentially at gas-liquid interfaces; late stages (t4–t5, forced oven evaporation) reveal transition to micron-sized aggregate formation following bulk brine pool depletion and loss of film continuity. C. Magnified view showing residual brine retention within porous salt at t5, even after oven drying, demonstrating how internal voids and inter-crystallite porosity sustain continued growth by maintaining brine access to active precipitation sites. D. Three-dimensional rendering of brine-salt coexistence around glass bead with continuous liquid film coating the bead surface, overlain by salt precipitates and internal inter-crystallite porosity. E. Segmented pore space at t5 showing banded aggregates preferentially obstructing pore throats and flow percolation pathways. F. Cross-sectional visualization revealing internal architecture with labeled structural features: ((1) umbrella-like surface crusts formed by progressive accumulation at gas-brine interface where evaporation is most intense, trapping residual brine within underlying pore spaces and prolonging drying; (2) internal aggregate structure composed of sub-micron halite crystal networks interspersed with voids and brine-coated surfaces; (3) fine-grained cubic halite crystals (1–10 µm) indicating localized growth under moderate supersaturation; (4) halite nucleating at brine-gas interface and growing into gas phase, forming voids between salt aggregates and bead surfaces as crystals preferentially expand into less viscous gas environment; (5) secondary porous networks formed through crystal aggregation on existing substrates with continuous brine supply, controlling fluid flow and ion transport pathways. G. Quantitative phase volumes evolution showing progressive salt accumulation (~10% at t5), transient brine volume increase during intermediate stages (t2–t3) reflecting capillary-driven solute delivery to precipitation sites, and final dry-out as water saturation approaches zero. Kernel density estimates with shaded regions representing ±5% confidence intervals from subvolume spatial variability. Glass beads occupy 60±2% of flow cell volume throughout all stages.

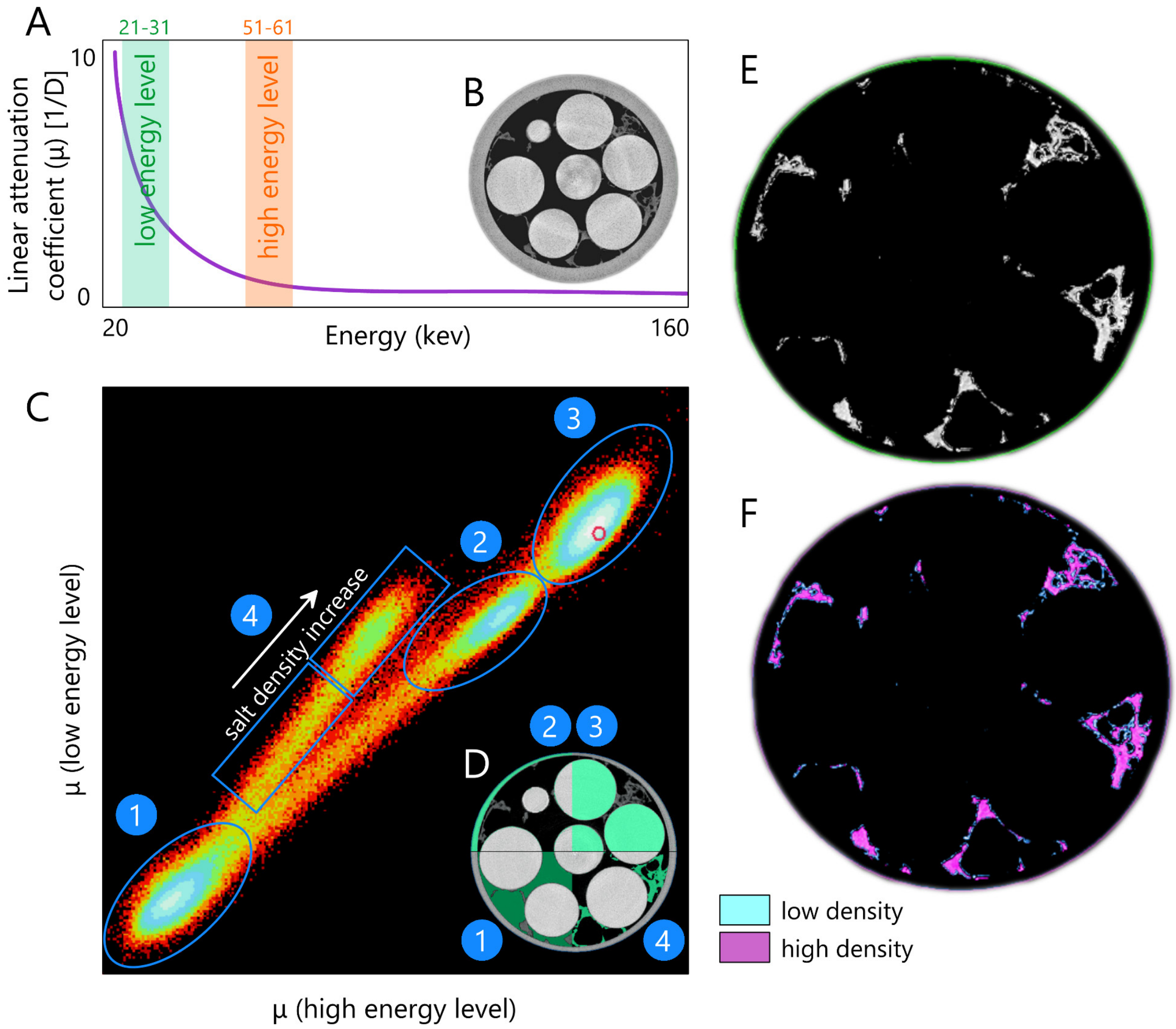

**Figure 3. Spectral tomography reveals bimodal density distribution within halite aggregates.** A. Schematic of the energy-dependency of the linear attenuation coefficient μ, with indication of the low- (21–31 keV) and high-energy (51–61 keV) ranges used to construct the multi-energy histogram of μ presented in C. B. Sp-CT image with spatial distribution of average μ values. C. Multi-energy histogram of μ obtained for the low energy range 21-31 keV versus the high energy range 51-61 keV, revealing four distinct material phases that can be associated with fluid (region 1), glass flow cell walls (region 2), glass beads (region 3), and halite aggregates (region 4). Within the halite cluster, μ values span from low-density porous precipitates to dense, compact aggregates, indicating substantial internal heterogeneity that is invisible to conventional energy micro-CT. D. Segmented phase distribution based on attenuation signatures from regions marked in panel C. E. Cross-sectional xy profile showing halite precipitation on glass bead surfaces and extending into adjacent pore spaces. F. Density-segmented profile revealing systematic bimodal architecture: dense macrocrystalline cores surrounded by lower-density porous microcrystalline overgrowths, particularly pronounced at top and bottom aggregate surfaces.

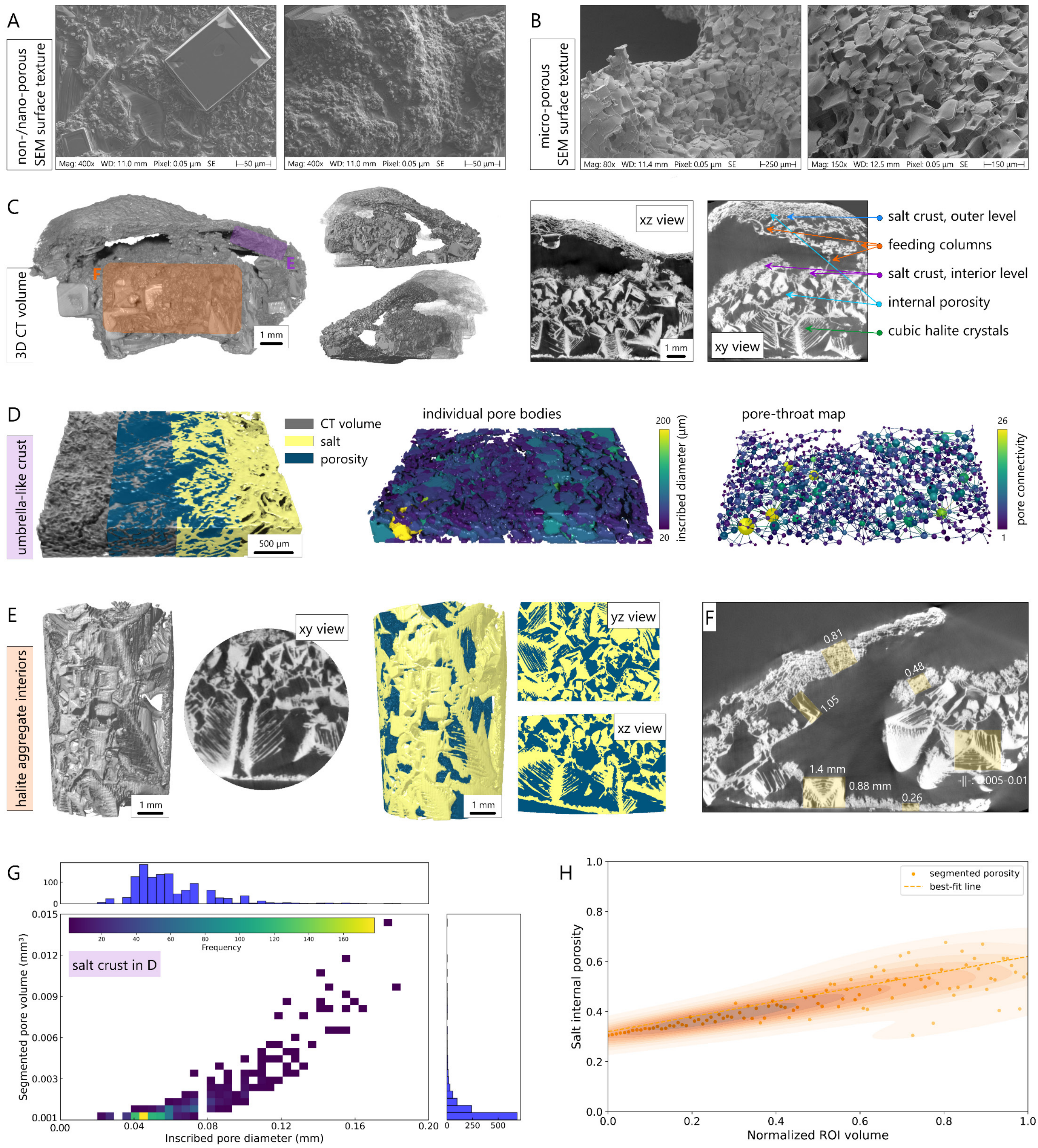

**Figure 4. Three-dimensional internal architecture and multiscale pore network characterization of halite aggregates.** A. Scanning electron microscopy (50 nm resolution) showing smooth, dense surface texture with minimal visible porosity (non-porous or nano-porous at resolution limit). B. SEM image revealing micro-porous texture with interconnected brick-like pore structures, indicating sustained pore connectivity during growth. C. Three-dimensional micro-CT renderings of extracted halite body showing precipitation texture, growth morphology, and multiple layered structures. Umbrella-like surface crust (color-coded violet) encapsulates the interior while growing upward into the gas phase. Cross-sectional profiles (xz, xy views) reveal intercrystalline porosity and feeding columns through which brine was transported via capillary forces to the evaporation-active top layer, where precipitation conditions were most favorable. Internal structure shows cubic halite crystals, internal voids, and porosity forming secondary substrates for the overlying crust layers. D. Pore network model derived from connected component analysis of surface crust, with pore bodies color-coded by inscribed diameter and pore-throat connectivity maps. Surface crusts contain approximately 40% internal porosity in three dimensions despite appearing non-porous in two-dimensional top-view sections. E-F. Extracted cylindrical region, cross-sectional profiles with corresponding segmentation showing halite aggregate interiors. Dimensional annotations (in millimeters) characterize the base layer thickness, fully developed euhedral to subhedral cubic crystals with columnar structures and lateral protrusions that create internal void spaces, as well as crosscutting crust layers that incorporate solute feeding columns. Internal porosity exhibits strong connectivity in both horizontal and vertical directions, establishing pervasive three-

dimensional networks that interconnect with both the basal and crustal layers. G. Plot of segmented pore volume against inscribed pore diameter in the extracted salt body, displaying the frequency of different pore size ranges. H. Internal porosity plot of extracted regions of interest within X-ray tomography volumes, showing increasing internal porosity and scatter with the increase of investigation volume. Best-fit trends and kernel density estimates indicate that larger analysis volumes capture greater structural heterogeneity, while smaller volumes sample more uniform local porosity, reflecting the hierarchical organization of pore networks within salt aggregates.